\documentclass[runningheads]{llncs}
\usepackage[T1]{fontenc}
\usepackage{graphicx}
\usepackage{todonotes}
\usepackage{hyperref}
\usepackage{color}

\renewcommand{\arraystretch}{1}
\usepackage[margin=1.2in]{geometry}

\usepackage{caption}
\usepackage{bbm}
\usepackage{algorithm}
\usepackage{algpseudocode}
\usepackage{amsmath}
\usepackage{comment}
\usepackage{tcolorbox}
\tcbuselibrary{skins,breakable}
\usepackage[utf8]{inputenc}
\usepackage{xcolor}
\usepackage{listings}
\usepackage{booktabs}
\usepackage{graphicx}
\usepackage{booktabs}
\usepackage{array}
\usepackage{xcolor}
\usepackage[table]{xcolor}
\usepackage{float}
\usepackage[multiple]{footmisc}
\usepackage{makecell}

\usepackage{fvextra}
\usepackage{pdflscape}
\begin{document}
\title{Portable Acceleration of Learning With Errors KEMs for Post-Quantum Cryptography}
%
%
\author{
    Tiziana Liberati\inst{1,3}\and
    Nitin Shukla\inst{2}\and
    Simone Rizzo\inst{1}\and
    Elisabetta Boella\inst{1}\and
    Matteo Barbieri\inst{1,4,5}\and
    Gabriella Bettonte\inst{1}\and
    Daniele Gregori\inst{1}\and
    Marco Pedicini\inst{3}
}
\authorrunning{T. Liberati et al.}
%
\institute{
E4 Computer Engineering SpA, via Martiri della Libertà 66, 42019 Scandiano (Italy) 
\email{\{tiziana.liberati, simone.rizzo, elisabetta.boella, matteo.barbieri, gabriella.bettonte,  daniele.gregori\}@e4company.com}
\and
SuperComputing Applications and Innovation Department, Cineca, via Magnanelli 6/3, 40033 Bologna (Italy)
\email{n.shukla@cineca.it}
\and
Dipartimento di Matematica e Fisica, Roma Tre University, Largo San Leonardo Murialdo 1, 00146 Rome (Italy)
\email{marco.pedicini@uniroma3.it}
\and
Dipartimento di Fisica e Astronomia “G. Galilei”, Universit\`a di Padova, via F. Marzolo 8, 35131 Padova (Italy)
\and INFN, Sezione di Padova, Italy
}

\maketitle              
\begin{abstract}

The transition to post-quantum cryptography (PQC) is driving demand for implementations that can meet the computational requirements of real-world applications. Among the proposed PQC constructions, Learning With Errors (LWE) based key encapsulation mechanisms (KEMs) are particularly attractive due to their strong security foundations, but they incur substantial computational costs from matrix operations and large-scale cryptographically secure random number generation. These characteristics position GPU acceleration as an effective approach for lowering the computational overhead of lattice-based cryptographic schemes.
In this work, we present a portable GPU implementation of a plain LWE-based KEM using OpenMP Target offloading. Unlike most existing GPU implementations, which rely on CUDA-specific optimizations, our approach uses a single source code base that executes on both NVIDIA and AMD accelerators.
To enable fully GPU-resident execution on heterogeneous platforms, we extend the \texttt{RNGonGPU} library with the Heterogeneous-computing Interface for Portability (HIP) support and integrate it into the OpenMP Target workflow.
We evaluate the proposed implementation on different accelerator architectures, analyzing performance benchmarking, runtime profiling, scalability analysis, and energy-to-solution measurements.
Experimental results show that OpenMP Target offloading delivers substantial acceleration over a multicore CPU baseline while preserving source-level portability across heterogeneous GPU ecosystems. Cross-platform analysis identifies NVIDIA GH200 and AMD MI300X as the most effective platforms for this memory-bound workload, while profiling indicates that memory-system organization and CPU–GPU interaction play a more critical role than peak compute capability alone. These findings demonstrate that portable GPU acceleration can significantly reduce the computational overhead of PQC while avoiding vendor lock-in, thereby facilitating the deployment of quantum-resistant cryptographic infrastructures.

\end{abstract}

\section{Introduction}
The transition to post-quantum cryptography (PQC) has intensified the demand for implementations that can meet the computational requirements of real-world applications. In this work, we focus on lattice-based cryptography~\cite{Regev}, one of the most widely studied families of post-quantum schemes, and in particular on Learning With Errors (LWE)-based key encapsulation mechanisms (KEMs). The computational cost of these schemes is largely driven by large-scale matrix operations and large-scale cryptographically secure
random number generation~\cite{KUMAR2022100242}. Since these workloads expose a high degree of parallelism, GPUs are a natural platform for accelerating LWE-based cryptographic schemes.
Previous GPU studies accelerated lattice-based PQC scheme such as FrodoKEM, NewHope, Kyber, and Dilithium~\cite{Dilithium,pqcgpu}. These implementations showed that GPU parallelism can provide high throughput for post-quantum cryptographic kernels, but they mainly relied on CUDA-oriented code paths and were therefore tied to the NVIDIA software ecosystem.
Within this literature, Kyber-oriented GPU implementations address a complementary point in the design space: Kyber relies on structured module-lattice polynomial arithmetic over a relatively small modulus, whereas the plain LWE KEM studied here is dominated by dense matrix-vector products and GPU-resident random sampling. Our motivation is therefore twofold: first, to quantify concrete acceleration for dense lattice-based PQC workloads; and second, to do so via a standards-based OpenMP Target/Heterogeneous-Compute Interface for Portability (HIP) path rather than a CUDA-specific application code path.
In previous work, we investigated the acceleration of an LWE-based KEM written in C++ using OpenACC directives~\cite{liberati}. While OpenACC offered a productive programming model and competitive performance on NVIDIA GPUs, its portability across heterogeneous accelerator ecosystems is limited. Standards-based models such as OpenMP Target provide broader support across vendor toolchains and therefore represent a promising approach for developing portable GPU-accelerated cryptographic software. This motivates our study of whether high-performance PQC implementations can be achieved while preserving source-level portability.
In this work, we investigate the portable acceleration of a plain LWE-based KEM using OpenMP Target offloading. To support GPU-resident random number generation on both AMD and NVIDIA accelerators, we extend the \texttt{RNGonGPU} library\footnote{\texttt{RNGonGPU} Git repository: \url{https://github.com/Alisah-Ozcan/RNGonGPU}} with HIP support. We then conduct a detailed benchmarking and profiling study to evaluate the runtime behavior and performance portability of OpenMP Target across NVIDIA (A100, GH200) and AMD (MI300X, MI300A) accelerators. Thus, the primary contributions of this study are:
\begin{itemize}
    \item A portable and open-source OpenMP Target implementation of a plain LWE-based KEM for both NVIDIA and AMD accelerators, enabling reproducible experiments and benchmarks.
    \item An extension of the \texttt{RNGonGPU} library with HIP support and its integration into the OpenMP Target workflow, enabling fully GPU-resident cryptographically secure random number generation.
    \item A comprehensive performance, energy-efficiency, and scalability study across different parameter sets, problem sizes, and heterogeneous GPU architectures.
\end{itemize}

The paper is organized as follows. Section~\ref{sec:math} introduces the mathematical foundations of the LWE-based KEM and describes the KEM workflow. Section~\ref{sec:impl} presents the GPU offloading strategy using OpenMP Target directives together with the integration of the \texttt{RNGonGPU} library. Section~\ref{sec:perf} discusses the experimental setup and performance evaluation, while Section~\ref{sec:profiling} presents a detailed runtime profiling analysis of OpenMP Target offloading on different GPU architectures. Finally, Section~\ref{sec:concl} concludes the paper and outlines future work.



All performance evaluations and benchmark reproductions in this work are based on the implementation hosted in the \emph{LWE-KEM-OMP-Target} Git repository\footnote{\hypertarget{fn:myfoot}{LWE-KEM Git repository: \url{ https://github.com/TizianaLiberati/LWE-KEM-OMP-Target}}}.

\section{Background and Scheme Overview}
\label{sec:math}
This section introduces the cryptographic setting and the LWE-based KEM used in our work, focusing on the aspects most relevant for parallelism and GPU offloading that will be presented in later sections.

\subsection{Mathematical Framework}

LWE-based cryptography derives its security from the hardness of solving noisy linear systems over finite fields:
\begin{equation}
    \label{eq:lwe}
    b = As+e \pmod q,
\end{equation}
where $A$ is a public matrix, $s$ is the secret vector, and $e$ is a small noise term. LWE is widely used in PQC because its security relies on hard lattice problems for which no efficient classical or quantum algorithms are currently known~\cite{Regev}. 

We now briefly describe the KEM workflow; for more details refer to the public repository of this work. A KEM is a protocol that allows two parties to agree on a shared secret key over an untrusted channel. The protocol uses a public-key encryption scheme as a subroutine, which consists of: \emph{KeyGen} (generate a public key $pk$ and a secret key $sk$); \emph{Encrypt} (encrypt a random message $m$ into a ciphertext); and \emph{Decrypt} (recover $m$ from the ciphertext).
The algorithms used within the KEM protocol are \emph{Encaps} (given $pk$, output a ciphertext and a shared key $K$, using the Encrypt routine); and \emph{Decaps} (given $sk$ and the ciphertext, recover the same $K$, using both Encrypt and Decrypt).

During decapsulation, we use the Fujisaki-Okamoto (FO) transform~\cite{FO}, which handles decryption failures in a cryptographically sound way and upgrades the overall security. Since both the CPU and GPU perform equivalent modular arithmetic and noise sampling, any decryption failure is attributable to the parameter choices rather than to hardware-specific effects.

\section{Implementation and Parallelization of the Post-Quantum Cryptosystem}
\label{sec:impl}

In this section, we analyze the main performance bottlenecks of the scheme and present our parallelization and offloading strategies.
The optimization process was initially developed for multicore CPUs, with OpenMP, and was later extended to GPU accelerators through OpenMP Target offloading. The discussion focuses primarily on the OpenMP Target design, data management, and GPU kernel organization.

\subsection{Performance Bottlenecks}
The main performance bottlenecks of the proposed plain LWE-based KEM arise from the repeated execution of dense matrix-vector multiplications under modular arithmetic, as defined in Equation~\eqref{eq:lwe}. These operations occur multiple times during key generations, encapsulation, and decapsulation, and dominate the overall computational cost of the protocol. Additional overhead arises from per-bit encapsulation/decapsulation and large-scale cryptographically secure random number generation. These operations are embarrassingly parallel and therefore well suited to GPU execution. 

As the lattice dimension $n$ increases, both computational complexity and memory traffic grow substantially because the public matrix scales as $O(n^2)$.

For large lattice dimensions, the public matrix $A$ frequently exceeds cache capacity, causing the workload to become strongly memory-bound.
Consequently, the performance of the implementation depends not only on arithmetic throughput, but also on memory bandwidth, data locality, and communication overhead between host and device memories.

Another major computational component is cryptographically secure random number generation. Post-quantum cryptographic schemes require large amounts of pseudorandom data derived from a secure seed and expanded through cryptographic primitives. To support GPU-resident random number generation, we integrate the \texttt{RNGonGPU} library~\cite{RNGonGPU}.

The original \texttt{RNGonGPU} implementation was designed for CUDA-based systems and relied on NVIDIA-specific runtime assumptions. Supporting AMD accelerators therefore required hipifying the CUDA backend, adding a HIP-compatible adapter layer, and integrating that adapter with OpenMP Target device data. In practice, the KEM kernels remain expressed with OpenMP directives, while the random-number backend is compiled through the HIP toolchain and is called on OpenMP-managed device buffers.
The interoperability mechanism used to expose these device buffers to the external HIP-based RNG backend relies on \texttt{\#pragma omp target data use\_device\_ptr}. This extension is important because it prevents random samples from being generated on the host and copied repeatedly to the accelerator; instead, randomness expansion, noise generation, and the dominant linear-algebra kernels remain GPU-resident on both NVIDIA and AMD systems.

\paragraph{Offloading Strategy.}
The OpenMP Target implementation counts on persistent device-side memory management and large-scale parallel execution of computational kernels. 

\paragraph{Device Residency and Memory Management.}
A primary optimization is the use of persistent device memory regions through \texttt{target enter data map(alloc:\ldots)} directives. All major matrices, ciphertexts, plaintexts, and temporary buffers are allocated once through persistent OpenMP Target data regions and remain resident on the GPU throughout execution. On NVIDIA systems, allocations are backed by CUDA Unified Memory through the \texttt{-gpu=managed} runtime. This strategy removes repeated allocation overheads from the critical execution path.
Data movement between host and device is managed explicitly through OpenMP \texttt{target update} directives and is performed only when CPU-side operations require access to GPU-resident data. 
The ciphertext equality check required by the FO transform is now performed entirely on the GPU, returning only a single integer flag to the host. Consequently, most stages of the KEM workflow remain resident on the device throughout execution while communication overhead is minimized.

\paragraph{GPU Kernel Parallelization.}
We now describe the main computational kernels parallelized with OpenMP Target directives, corresponding to the key generation, encryption, and decryption stages that dominate the execution time of the LWE-based KEM workflow.

\textbf{Key Generation.}
Key generation is implemented through three GPU-resident phases, as illustrated in Listing~\ref{lst:omp-keygen}. First, uniformly distributed random values are generated on the device and converted in parallel into the secret vector $s$. Second, Gaussian noise samples are generated on the GPU, clamped, and rounded to obtain the integer error vector $e$. Finally, the public key vector $t = A s + e \bmod q$ is computed through a dense matrix--vector product distributed across GPU teams. This organization keeps both randomness generation and linear algebra operations GPU-resident. 

\begin{lstlisting}[language=C++, caption={GPU-parallel key generation structure.},
label={lst:omp-keygen}]
/* Phase 1: ternary secret vector */
rngongpu_fill_uniform_u32(sraw, key_seed);

#pragma omp target teams distribute parallel for
for (i = 0; i < n; ++i)
    s[i] = ternary_sample(sraw[]);

/* Phase 2: Gaussian error vector */
rngongpu_fill_normal(etmp, key_seed);

#pragma omp target teams distribute parallel for
for (i = 0; i < n; ++i)
    e[i] = round_and_clamp(etmp[]);

/* Phase 3: public key vector */
#pragma omp target teams distribute parallel for
for (i = 0; i < n; ++i)
    t[i] = dot(A[i,*], s) + e[i] mod q;
\end{lstlisting}

\textbf{Encryption and Decryption.}
Encryption is parallelized across independent message bits, as shown in Listing~\ref{lst:omp-encrypt}. All per-bit encryptions are fused into a single batched GPU kernel: each GPU team is responsible for one ciphertext component, while inner products over lattice dimension $n$ are parallelized within the team. This structure reduces kernel-launch overhead, improves occupancy, and maximizes the reuse of device-resident buffers.

\begin{lstlisting}[language=C++, caption={Batched GPU encryption kernel.},
label={lst:omp-encrypt}]
#pragma omp target teams distribute
for (j = 0; j < msg_bits; ++j) {

    #pragma omp parallel for simd
    for (i = 0; i < n; ++i)
        u[j][i] = dot(A[:,i], r[j]) + e1[j][i] mod q;

    #pragma omp parallel for simd reduction(+:dot)
    for (i = 0; i < n; ++i)
        dot += t[i] * r[j][i];

    v[j] = dot + e2[j] + m[j] mod q;
}
\end{lstlisting}
The \texttt{simd} clauses in Listing~\ref{lst:omp-encrypt} expose the innermost reductions and element-wise ciphertext updates to the compiler, enabling vectorized code generation for the memory-streaming loops. In our measurements, their impact on overall performance was modest and compiler dependent. The dominant performance gains arise from batching 256 independent encryptions, keeping buffers resident on the device, and exploiting the large degree of thread-level parallelism available on the GPU.
Nevertheless, we retain the \texttt{simd} clauses because they explicitly express the absence of loop-carried dependencies and improve portability across OpenMP implementations.

Decryption follows an analogous batched structure. Each GPU team processes one ciphertext, computes the dot product between the secret key $s$ and the first ciphertext component $c$, and reconstructs the encoded plaintext coordinate by evaluating $\mu = v - \langle s, u\rangle \pmod q$.
A threshold decision rule is then applied to recover the corresponding message bit.

Additional GPU parallelization is employed during ciphertext verification, where the FO equality check is implemented as an on-device OpenMP reduction. Ciphertext buffers are compared directly on the GPU, and only a single mismatch flag is returned to the host, avoiding unnecessary device-to-host transfers.

Additional GPU-side parallelism is also exploited during random number generation through the integration of the \texttt{RNGonGPU} library, which provides an AES-256~\cite{aes} Deterministic Random Bit Generator compliant with NIST SP 800-90A~\cite{NISTSP80090A}. Only a small initialization seed is transferred from the CPU, while random expansion remains GPU-resident. This significantly reduces communication overhead and allows most stages of the KEM workflow to remain resident on the device.

\section{Performance evaluation and scalability tests}
\label{sec:perf}
This section evaluates the performance and scalability of the proposed OpenMP Target implementation across heterogeneous GPU architectures. We first describe the experimental platforms and benchmarking methodology, then compare OpenMP multicore execution against GPU offloading, and finally analyze performance, portability, and runtime behavior through cross-platform profiling.

\subsection{Hardware Configuration \& Methodology}\label{sec:HW}

One of the goals of this work is to evaluate the performance portability of OpenMP Target across heterogeneous accelerator architectures. To this end, the same OpenMP Target implementation is executed without architecture-specific code modifications on four different GPU platforms spanning both NVIDIA and AMD ecosystems.
The executables for NVIDIA systems are compiled using the NVIDIA HPC SDK (NVHPC 25.7), while the ones for AMD platforms are compiled using the ROCm 7.2 LLVM/Clang toolchain. This allows the same source code to be evaluated through the vendor-supported OpenMP offloading implementations available on each architecture. 
The evaluated accelerators include NVIDIA A100-SXM-64GB, NVIDIA GH200 Grace Hopper, AMD MI300X, and AMD MI300A. Table~\ref{tab:gpu_architecture} reports a selection of hardware characteristics that are relevant to this work. The A100 provides 64\,GB of HBM2e memory with 1640\,GB/s peak bandwidth and represents a conventional PCIe-attached discrete accelerator. The GH200 integrates a Hopper GPU with a Grace CPU connected through NVLink-C2C, exposing 96\,GB of HBM3 memory with approximately 4\,TB/s bandwidth and significantly reducing CPU--GPU communication overhead. The AMD MI300X is a high-end discrete accelerator equipped with 192\,GB of HBM3 memory delivering up to 5.3\,TB/s bandwidth. Finally, the MI300A adopts an APU design in which CPU and GPU chiplets share a unified 128\,GB HBM3 memory pool via AMD coherent Infinity Fabric, eliminating the traditional distinction between host and device memory.

\begin{table}
\caption{Key architectural characteristics of the evaluated accelerators.}
\label{tab:gpu_architecture}

\centering
\normalsize

\setlength{\tabcolsep}{4pt}
\renewcommand{\arraystretch}{0.95}

\resizebox{\linewidth}{!}{
\begin{tabular}{
l
c c c c
}

\toprule

\textbf{Characteristic} 
& \textbf{NVIDIA A100} 
& \textbf{NVIDIA GH200}
& \textbf{AMD MI300A}
& \textbf{AMD MI300X}
\\

\midrule
Max clock frequency    & 1419\,MHz            & 1980\,MHz   & 2100\,MHz                & 2100\,MHz\\
Total number of cores  & 6912                 & 16896       & 14592           & 19456 \\
Peak FP64              &  9.7\,TFLOPS          & 34\,TFLOPS            & 61.3\,TFLOPS                & 81.7\,TFLOPS  \\
HBM type               & HBM2e                & HBM3        & HBM3            & HBM3\\
HBM bandwidth          & 1640\,GB/s           & 4\,TB/s     & 5.3\,TB/s       & 5.3\,TB/s \\
HBM capacity           & 64\,GB               & 96\,GB      & 128\,GB         & 192\,GB \\
CPU--GPU interconnect  & PCIe~4.0 $\times$16  & NVLink-C2C  & Infinity Fabric & PCIe~5.0 $\times$16\\

\bottomrule

\end{tabular}%
}
\end{table}


Unless otherwise specified, reported results correspond to steady-state execution and are obtained averaging over multiple simulation repetitions (typically ten) to account for run-to-run variability. 
Execution times are collected using \verb|std::chrono::steady_clock|, with the timing instrumentation integrated directly into the source code.

\subsection{OpenMP Multicore versus OpenMP Target Offloading}

Table~\ref{tab:performance_openomp_target} reports the average execution times of the multicore OpenMP CPU implementation and the OpenMP Target GPU implementation on NVIDIA GH200 for $N=100$ benchmark iterations, while varying the lattice dimension $n$. The results show that GPU offloading becomes increasingly advantageous as the lattice dimension grows, eventually achieving speedups exceeding two orders of magnitude for large problem sizes. As $n$ increases, the execution time is progressively dominated by \emph{Encaps} and \emph{Decaps} phases, indicating that the workload becomes strongly memory-bound due to repeated streaming accesses over the matrix $A$.

The large speedups should not be interpreted as a direct ratio between peak CPU and GPU specifications. The multicore CPU baseline is limited by a combination of memory bandwidth, cache capacity, synchronization, and the cost of repeatedly traversing the dense matrix and generating cryptographic randomness. For the largest tested dimension, the matrix footprint exceeds cache capacity and the CPU implementation spends most of its time in bandwidth-limited matrix-vector operation. The GPU implementation, in contrast, exposes both the matrix rows and the 256 independent ciphertext slots to massive thread-level parallelism, keeps the main buffers resident on the device, and amortizes kernel-launch and data-management overhead across batched operations. The resulting speedup therefore reflects both architectural bandwidth/parallelism and the more favorable execution organization enabled by OpenMP Target offloading.

The earlier OpenACC implementation~\cite{liberati} serves as an important reference point for the optimization strategy, since it demonstrated that this LWE-KEM workflow benefits substantially from accelerator execution on NVIDIA systems. However, it does not provide a cross-vendor baseline because the implementation and random-number backend were tied to the NVIDIA-oriented OpenACC/CUDA software stack. For this reason, the main quantitative comparison in this paper uses the multicore OpenMP implementation as the portable CPU baseline and evaluates the OpenMP Target version across both NVIDIA and AMD GPUs. This emphasizes the central question of the present study: whether a single OpenMP Target code base can preserve acceleration while extending execution to heterogeneous accelerator ecosystems.
\begin{table}
\caption{Performance comparison between the multicore OpenMP implementation (72 threads) and the OpenMP Target implementation on the GH200 for $N = 100$ iterations. The table reports the average execution times of the \emph{KeyGen}, \emph{Encaps}, and \emph{Decaps} phases (in $\mathrm{ms}$), as well as the total workflow runtime (in $\mathrm{s}$). The performance gain of OpenMP Target over OpenMP is quantified as $\mathrm{Speedup} = T_{\mathrm{OpenMP}} / T_{\mathrm{OpenMP Target}}$, which increases with $n$.}
\centering
\normalsize
\setlength{\tabcolsep}{3pt}
\renewcommand{\arraystretch}{1.0}

\resizebox{\linewidth}{!}{%
\begin{tabular}{
c 
c c c c c c c c 
c}

\toprule
& \multicolumn{4}{c}{\textbf{OpenMP}} 
& \multicolumn{4}{c}{\textbf{OpenMP TARGET}} 
& \textbf{} \\

\cmidrule(lr){2-5}
\cmidrule(lr){6-9}

\textbf{$n$}
& \textbf{KeyGen} & \textbf{Encaps} & \textbf{Decaps} & \textbf{Total}
& \textbf{KeyGen} & \textbf{Encaps} & \textbf{Decaps} & \textbf{Total}
& \textbf{Speedup} \\

\midrule
32 & 0.85 & 0.26 & 0.14 & 0.13 & 3.39&	5.79 & 5.6&	1.48 &  0.09\\
128  & 7.54	& 0.65 &0.58 &0.88 &  3.23 & 6.02 &5.63 & 1.49 & 0.59\\
256  &30.31 & 1.44 &1.34 & 3.31 & 3.24 & 6.35 & 5.66 & 1.53 & 2.17\\
512  & 118.68 & 3.91 & 3.79 & 12.65 & 3.31 & 7.49 & 6.31 &1.72 & 7.38\\
1024 & 455.85 & 12.46	& 12.24 & 48.08 & 3.64	& 8.63 & 6.22	& 1.85 & 25.95\\
2048 & 1820.44 & 45.61 & 45.65 & 191.25 & 4.59 & 13.94 & 9.85 & 2.84 & 67.27\\
4096 & 7227.88 &	173.39 & 173.59 & 757.89 & 8.48 & 31.81	& 22.81	& 6.31 & 120.02\\
\bottomrule
\end{tabular}%
}
\label{tab:performance_openomp_target}
\end{table}


\subsection{Performance Portability Across GPU Architectures}


In this section, we evaluate the four accelerator platforms described in Section~\ref{sec:HW} to characterize the performance of the OpenMP Target implementation of the LWE-KEM benchmark across heterogeneous GPU architectures.
To this end, we perform a series of experiments with $N = 100$, varying the lattice dimension $n$ from 32 to 16384.
Execution times for the largest value of $n$ tested are listed in Table~\ref{tab:gpu_relative_slowdown}, which also reports the normalized slowdown relative to the GH200 baseline. The normalized slowdown is defined as: $\mathrm{Slowdown} = T_{\mathrm{GPU}}/T_{\mathrm{GH200}}$, where $T_{\mathrm{GPU}}$ denotes the execution time on a given accelerator and $T_{\mathrm{GH200}}$ the execution time on the NVIDIA SuperChip. The results show that at large values of $n$, GH200 provides the fastest execution times. The other platforms exhibit progressively larger performance degradation with respect to the GH200 baseline.

\begin{table}
\caption{Relative OpenMP Target performance across GPU platforms for $N=100$ and $n=16384$.}
\centering
\normalsize

\setlength{\tabcolsep}{4pt}
\renewcommand{\arraystretch}{0.95}

\resizebox{\linewidth}{!}{%
\begin{tabular}{
l
c c c
}

\toprule

\textbf{GPU Platform}
& \textbf{Execution Time [s]}
& \textbf{Relative Slowdown}
& \textbf{Performance Class} \\

\midrule

NVIDIA GH200 & 60.03 & $1.00 \times$ & Baseline \\
AMD MI300X   & 84.87  & $1.41 \times$ & Near-optimal \\
NVIDIA A100  & 96.83  &  $ 1.61 \times$ & Moderate slowdown \\
AMD MI300A   & 114.55 & $1.91\times$ & Largest slowdown \\

\bottomrule
\end{tabular}%
}

\label{tab:gpu_relative_slowdown}
\end{table}

\begin{figure}
    \centering
    \includegraphics[width=0.5\linewidth]{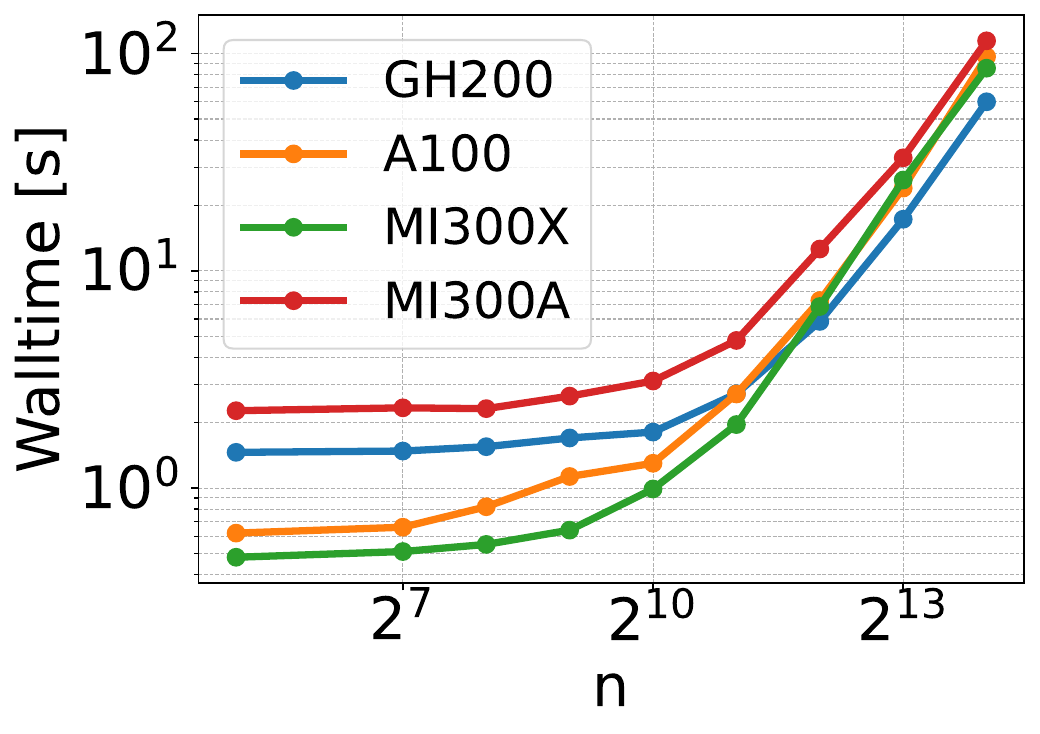}
    \caption{Scalability of the OpenMP Target implementation across different GPU architectures: GH200 (blue), A100 (orange), MI300X (green), and MI300A (red) for $N=100$ and increasing lattice dimension $n$. Total runtime is reported in seconds.}
    \label{fig:gpu_perf}
\end{figure}

Figure~\ref{fig:gpu_perf} reports the total execution time of the LWE-KEM versus $n$ across the evaluated GPU architectures. As the lattice dimension increases, the workload exhibits progressively stronger memory-bound behavior, with overall runtime increasingly dominated by the Encapsulation and Decapsulation phases.



Among the evaluated systems, the AMD MI300X and NVIDIA GH200 exhibit the lowest execution times for large lattice dimensions, which are the most relevant from a security standpoint. The MI300X provides the best performance for problem sizes up to $n = 4096$, whereas the GH200 becomes consistently faster as $n$ increases beyond this point. The strong performance of both accelerators is likely attributable to the high sustained bandwidth of their HBM3/HBM3e memory subsystems, whose importance grows as the workload becomes increasingly memory-bound.

For small lattice sizes, the NVIDIA A100 achieves execution times comparable to those of the MI300X. However, its performance deteriorates for larger problem sizes, consistent with the lower bandwidth of its HBM2e memory subsystem and its reduced ability to sustain the high streaming throughput required by memory-intensive workloads.

In contrast, the AMD MI300A exhibits the highest execution times and the weakest scalability across the evaluated configurations. This performance gap becomes increasingly pronounced for $n > 4096$, despite the availability of a high-bandwidth unified HBM3 memory subsystem. As discussed previously, this behavior is consistent with contention within the shared memory hierarchy: CPU-side cryptographic operations, including SHA3 hashing and the Fujisaki--Okamoto (FO) transform, execute concurrently with GPU kernels and compete for the same HBM3 memory resources. The resulting reduction in effective memory bandwidth limits the throughput of the memory-streaming kernels and ultimately degrades overall application performance.

We also evaluate performance in terms of energy-to-solution for the two best-performing GPUs, the NVIDIA GH200 and AMD MI300X. For this analysis, we consider two problem sizes: the first with $n = 16384$ and $N = 100$  and the second with $n = 32768$ and $N = 1000$.
We focus exclusively on the energy consumed by the GPU, neglecting the contribution of the remaining system components.

GPU energy consumption is estimated by numerically integrating the measured power draw over the execution time. Power measurements are collected at an approximate sampling frequency of $1\,\mathrm{Hz}$ using \verb|nvidia-smi| on the GH200 and \verb|amd-smi| on the MI300X~\cite{boella,Amati}. Table~\ref{tab:energy} reports the average execution time and GPU energy consumption per simulation. The results indicate that, for these large lattice sizes, the NVIDIA GH200 not only achieves lower execution times, but is also approximately $2.5\times$ more energy-efficient than the AMD MI300X.

We also observe that the execution times reported in Table~\ref{tab:energy} are slightly higher than those presented in Table~\ref{tab:gpu_relative_slowdown}. This difference highlights the overhead introduced by the power-monitoring procedure, which periodically samples the GPU power consumption during execution.

\begin{table}
\caption{Average Time-to-solution and Energy-to-solution for two selected cases.}

\centering
\normalsize

\setlength{\tabcolsep}{4pt}
\renewcommand{\arraystretch}{0.95}

\resizebox{\linewidth}{!}{%
\begin{tabular}{l c c c c}

\toprule

& \multicolumn{2}{c}{\textbf{$N=100$, $n=16384$}}
& \multicolumn{2}{c}{\textbf{$N=1000$, $n=32768$}} \\

\cmidrule(lr){2-3} \cmidrule(lr){4-5}

\textbf{GPU Platform}
& \textbf{Time-to-solution [s]}
& \textbf{Energy-to-solution [kJ]}
& \textbf{Time-to-solution [s]}
& \textbf{Energy-to-solution [kJ]} \\

\midrule

NVIDIA GH200 & $60.04 \pm 0.01$ & $9.70 \pm 0.07$ & $2362.42 \pm 0.27$ & $397.78 \pm 0.32$ \\
AMD MI300X & $85.80 \pm 0.49$ & $26.25 \pm 0.12$ & $3179.64 \pm 1.20 $ & $1030.58 \pm 3.43$ \\

\bottomrule

\end{tabular}%
}

\label{tab:energy}
\end{table}

\begin{figure}
    \centering
    \includegraphics[width=0.7\linewidth]{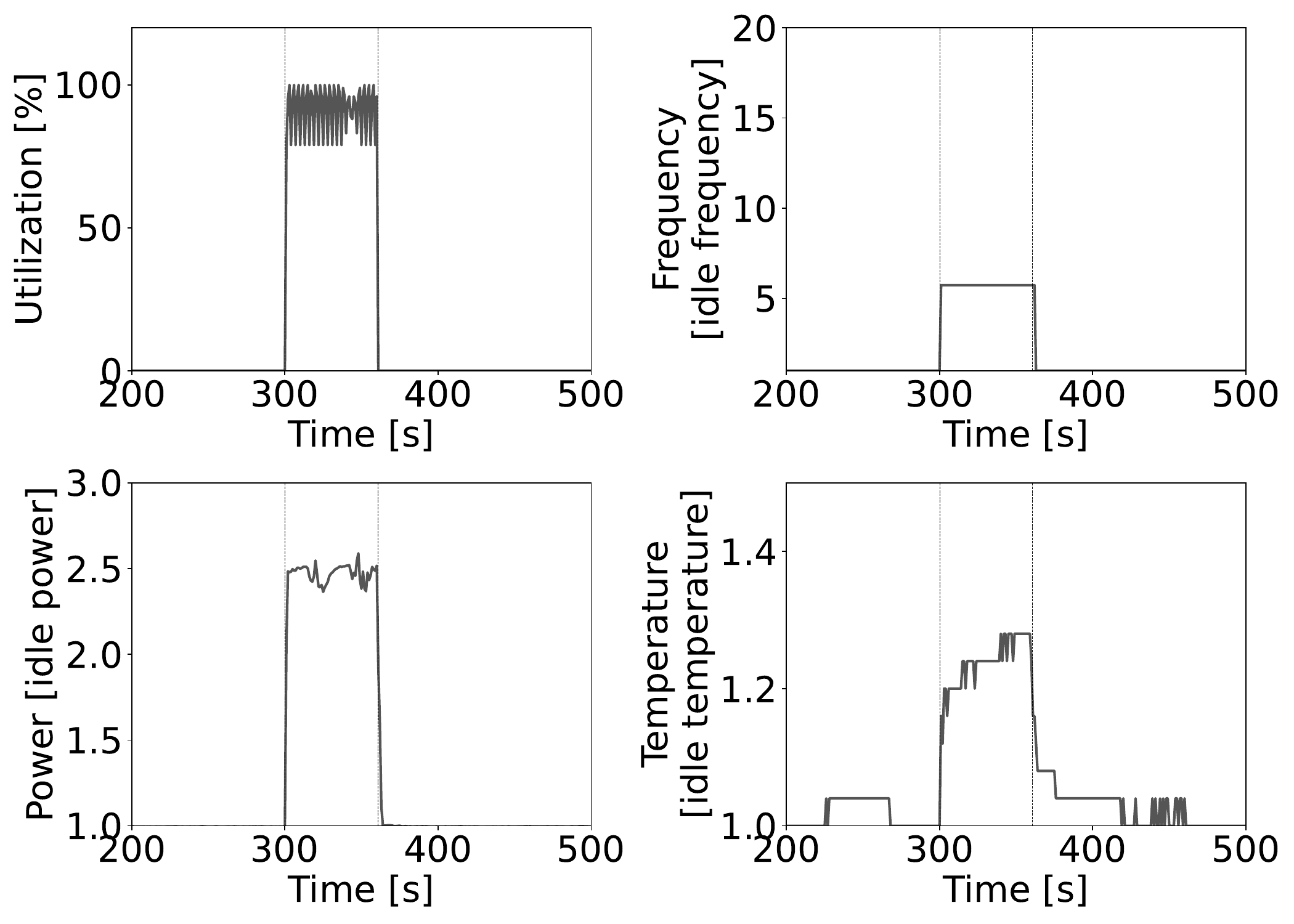}
    \caption{Evolution of key GPU metrics during a representative execution of the benchmark with $N=100$ and $n=16384$ on the GH200. The top-left, top-right, bottom-left, and bottom-right panels report GPU utilization, operating frequency, power consumption, and temperature, respectively. The beginning and end of the simulation are marked by black dashed lines. All quantities are normalized with respect to their corresponding idle-state values.}
    \label{fig:smi_GH200}
\end{figure}

\begin{figure}
    \centering
    \includegraphics[width=0.7\linewidth]{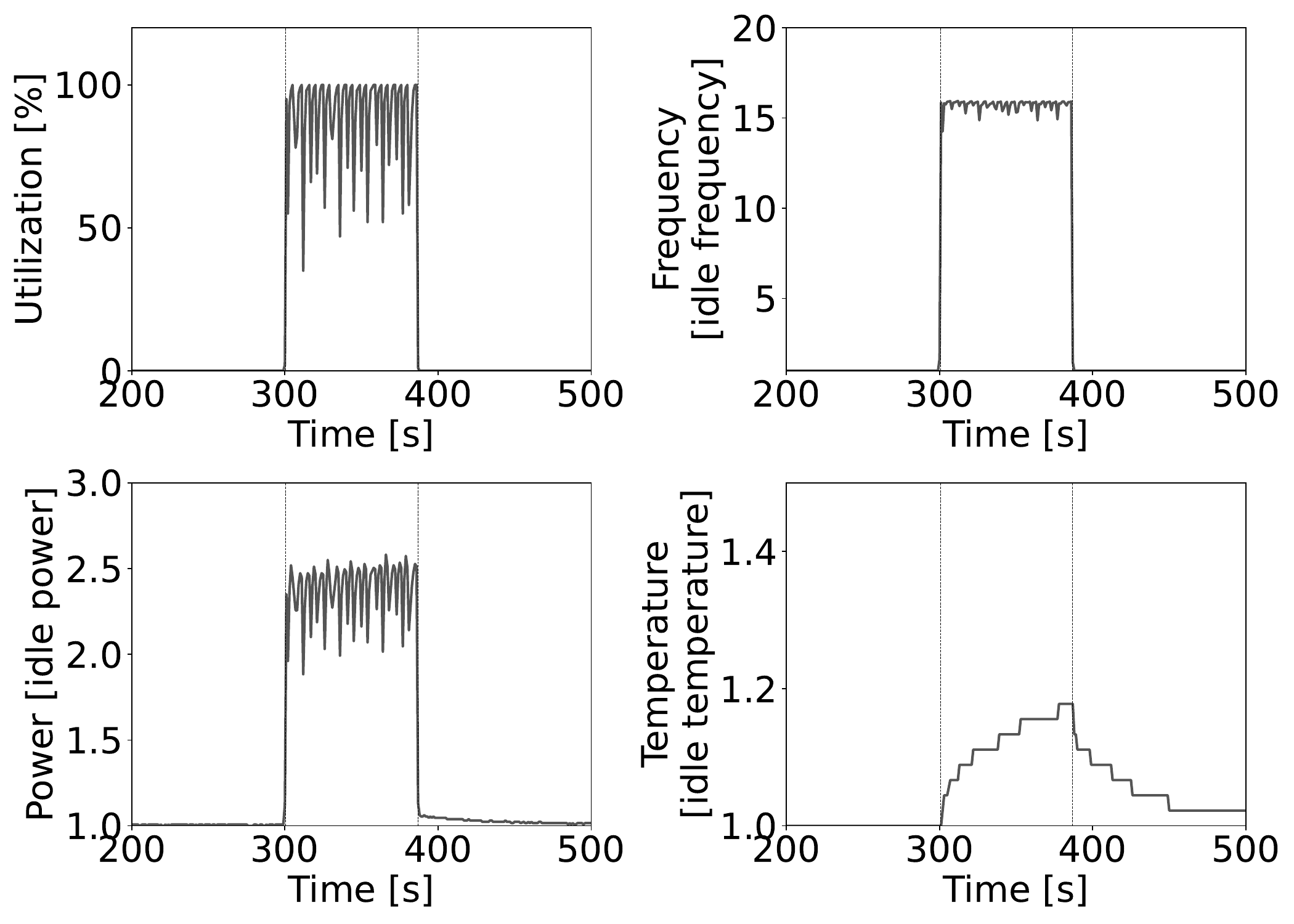}
    \caption{Equivalent to Figure~\ref{fig:smi_GH200}, showing the results for a run on the MI300X.}
    \label{fig:smi_MI300}
\end{figure}

To gain further insight into these results, we analyze the evolution of GPU utilization, operating frequency, power consumption, and temperature during a representative execution. These quantities, normalized to their respective idle-state values, are shown in Figures~\ref{fig:smi_GH200} and~\ref{fig:smi_MI300} for GH200 and MI300X, respectively.

During a typical run, the GH200 reaches peak utilization levels of 100\%, while maintaining occupancy above 80\% throughout most of the execution. Although the MI300X also reaches 100\% utilization, its utilization profile exhibits more pronounced fluctuations and frequent drops. At the beginning of the simulation, the GH200 frequency rapidly increases to its maximum operating value of $1980 \,\mathrm{MHz}$, approximately five times higher than the idle frequency, and remains essentially constant until the end of the run. In contrast, the MI300X frequency rises from an idle value of $132 \,\mathrm{MHz}$ to its maximum operating frequency of $2100 \,\mathrm{MHz}$, nearly a fifteen-fold increase, but subsequently exhibits small oscillations around this value.

In both systems, the power consumption is not constant during execution and displays an oscillatory behavior. The MI300X exhibits higher-frequency power fluctuations that correlate with the oscillations observed in the utilization profile. Furthermore, its power consumption shows a gradual upward trend over the course of the simulation, a behavior that is not apparent on the GH200. 
In absolute terms, the NVIDIA GH200 draws significantly less power than the MI300X (approximately $160 \,\mathrm{W}$ versus $300 \,\mathrm{W}$ on average) during the simulation. However, when normalized to the respective idle-state values, the increase in power consumption is comparable between the two systems.
The GH200 energy-to-solution advantage stems from both its shorter execution times and, more importantly, its lower power consumption.


The GPU temperature increases throughout the simulation on both platforms. By the end of the run, the temperature of the GH200 is approximately $1.3\times$ higher than its initial value, whereas the MI300X temperature increases by about $1.2\times$.

The utilization drops observed on the MI300X, accompanied by correlated fluctuations in power consumption, point to less efficient resource utilization and may reflect periods of memory- or runtime-induced stalls. Such behavior could partially explain the longer execution times observed on this platform. Conversely, no clear evidence of thermal or frequency throttling is apparent, as both GPUs remain close to their maximum operating frequencies throughout the run.

\section{Architecture-Driven Performance Divergence}
\label{sec:profiling}
This section explains the performance differences observed in Section~\ref{sec:perf} from an architectural and OpenMP runtime perspective. We analyze how memory bandwidth, host--device communication, unified memory, and compiler/runtime behavior influence the performance of the OpenMP Target implementation across the evaluated GPU architectures.

Execution traces were collected using NVIDIA Nsight Systems on the A100 and GH200 platforms, and ROCm Systems Profiler on the MI300X and MI300A platforms. Representative first-iteration timelines are shown in Figures~\ref{fig:timeline-gh200} and~\ref{fig:timeline-mi300a}.

\subsection{Cross-Platform Profiling Analysis}

Across all systems, the execution structure is broadly identical. Each KEM iteration consists of three principal phases: \textit{KeyGen}, \textit{Encaps}, and \textit{Decaps}. The dominant GPU activities correspond to matrix generation, key generation, batched encryption, and batched decryption. Among these, the Encapsulation stage dominates execution cost due to repeated large-scale accesses to the matrix $A \in \mathbbm{Z}_q^{n \times n}$ and the generation of all ciphertext slots in parallel.

Profiling results indicate that the benchmark is strongly memory-bound.
The dominant kernels stream continuously through the flattened matrix representation and associated noise buffers, making effective memory bandwidth the primary performance-limiting factor rather than peak computational throughput.

We now examine the profiling behavior of each GPU architecture.

\paragraph{NVIDIA A100-SXM-64GB.}
\label{sec:prof-a100}

Profiling results on the A100 indicate that execution time is dominated by memory-intensive operations associated with the encryption and decryption stages. The observed behavior is consistent with a memory-bound workload, where performance is constrained primarily by the available HBM2e bandwidth. Since CPU-side cryptographic operations execute on a separate DDR memory subsystem, no measurable CPU--GPU memory contention is observed. Consequently, performance is mainly determined by device-memory throughput and PCIe transfer overhead.

\paragraph{NVIDIA GH200 Grace Hopper SuperChip.}
\label{sec:prof-gh200}

\begin{figure}
    \centering
    \includegraphics[width=\linewidth]{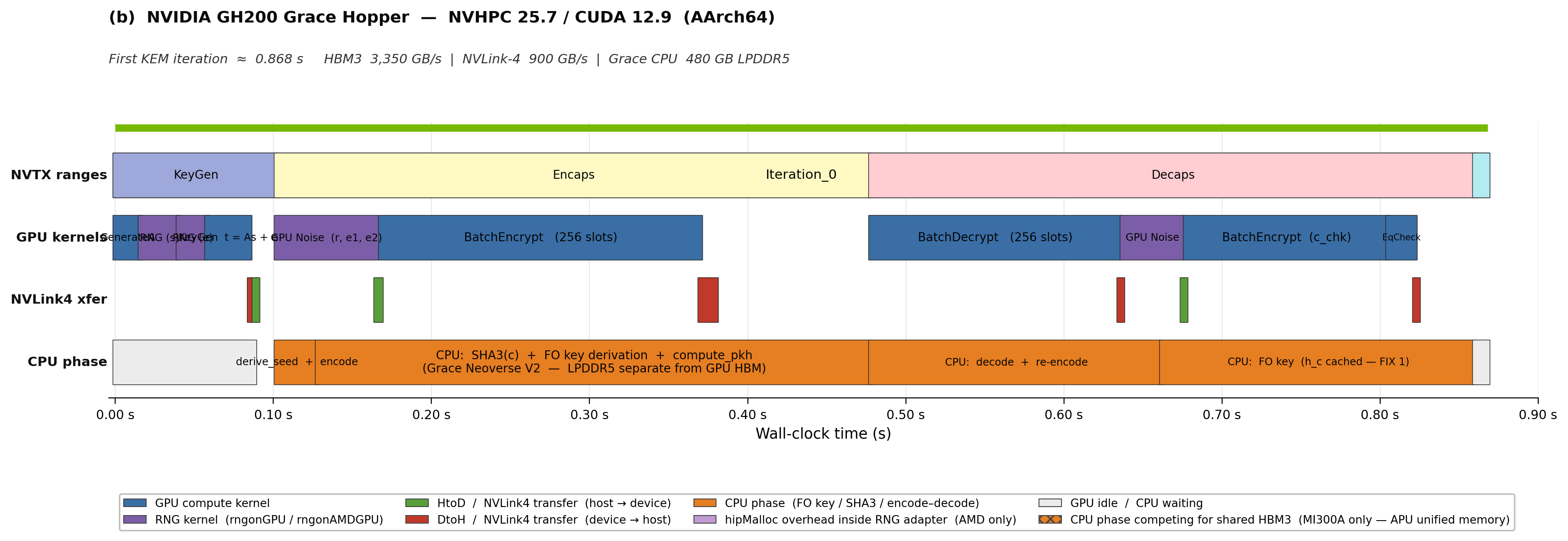}
    \caption{NVIDIA Nsight Systems timeline for the NVIDIA Grace Hopper SuperChip. NVLink-C2C substantially reduces ciphertext transfer latency relative to PCIe-based systems. The 4.1\,MB ciphertext transfer decreases from approximately 51\,ms on the A100 to approximately 9\,ms on the GH200. CPU-side operations execute on a separate LPDDR5 subsystem, avoiding contention with GPU HBM3 accesses.}
    \label{fig:timeline-gh200}
\end{figure}

The profiling timeline for the GH200 is shown in Figure~\ref{fig:timeline-gh200}. Although the overall execution pattern remains similar to that observed on the A100, the GH200 benefits from both higher HBM3 bandwidth and the NVLink-C2C interconnect between the Grace CPU and Hopper GPU. These architectural features reduce both kernel execution times and communication overheads: the 4.1\,MB ciphertext transfer decreases from approximately 51\,ms on the PCIe~4.0 A100 system to approximately 9\,ms on the GH200. An important architectural characteristic of the GH200 is the physical separation between Grace LPDDR5 memory and Hopper HBM3 memory. As a result, CPU-side SHA3 hashing and FO key derivation do not consume GPU memory bandwidth, and no measurable CPU--GPU memory contention is observed. The GH200 therefore combines efficient host--device communication with the isolation of CPU and GPU memory traffic, contributing to its near-optimal performance for this memory-bound workload.

\paragraph{AMD MI300X.}
\label{sec:prof-mi300x}

The MI300X achieves execution times consistent with the high sustained throughput of its HBM3 memory subsystem, which is particularly beneficial for the strongly memory-bound nature of the LWE-KEM workload. Compared with PCIe~4.0 systems, the PCIe~5.0 interconnect further reduces communication overhead, although host--device transfers remain more expensive than on the GH200 equipped with NVLink-C2C.
Profiling also reveals a runtime overhead associated with GPU-resident random number generation on AMD platforms. Dynamic memory allocations performed within the HIP runtime introduce additional latency that is absent on the NVIDIA platforms and does not originate from the cryptographic computation itself. This overhead identifies an implementation limitation at the OpenMP-HIP interoperability boundary: OpenMP manages the lifetime of the KEM buffers, but allocations internal to the external HIP RNG backend are still governed by HIP runtime behavior. Persistent pre-allocation inside the RNG adapter would likely reduce this overhead.

\paragraph{AMD MI300A.}
\label{sec:prof-mi300a}

\begin{figure}
    \centering
    \includegraphics[width=\linewidth]{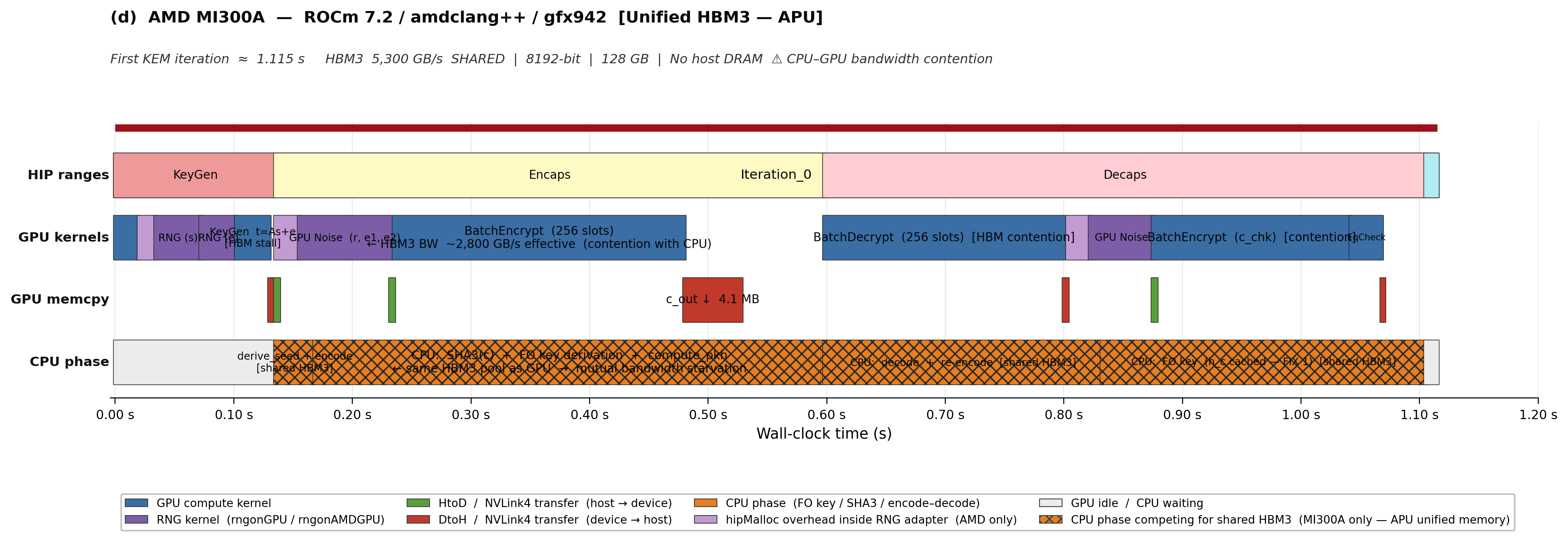}
    \caption{ROCm Systems Profiler timeline for the AMD MI300A. CPU-side SHA3 hashing and FO key derivation overlap with GPU execution while sharing the same HBM3 memory subsystem. The resulting CPU--GPU memory contention reduces the effective memory bandwidth available to the workload and contributes to the longer execution times observed relative to the discrete MI300X configuration.}
    \label{fig:timeline-mi300a}
\end{figure}

Despite its high-bandwidth unified-memory design, the MI300A exhibits the slowest execution time among the evaluated systems. The profiling timeline shown in Figure~\ref{fig:timeline-mi300a} highlights the main architectural difference with respect to discrete accelerators: CPU and GPU accesses are serviced from the same HBM3 memory pool and memory-controller infrastructure.

CPU-side SHA3 hashing and FO key derivation overlap with GPU execution during both the encapsulation and decapsulation phases. As a result, CPU memory accesses and GPU streaming operations contend for the same HBM3 memory resources, reducing the effective bandwidth available to the accelerator. Profiling reveals a substantial increase in the execution time of the dominant memory-bound kernels compared with the MI300X. Specifically, the encryption phase increases from approximately $177\,\mathrm{ms}$ on the MI300X to $245\,\mathrm{ms}$ on the MI300A, while the decryption phase grows from approximately $146\,\mathrm{ms}$ to $202\,\mathrm{ms}$, corresponding to a consistent slowdown of approximately $1.38\times$ for both kernels.

Since the MI300A and MI300X share the same GPU architecture, ROCm software stack, and kernel implementation, the observed degradation cannot reasonably be attributed to differences in computational throughput, compiler optimizations, or software behavior. Instead, it is consistent with contention within the shared HBM3 memory subsystem. On the MI300A, CPU-side SHA3 hashing and FO key derivation execute concurrently with GPU streaming accesses, issuing competing requests to the same physical HBM3 pool and thereby reducing the effective memory bandwidth available to the GPU kernels.

These results highlight an important limitation of unified CPU--GPU memory architectures for memory-bound cryptographic workloads. Although the MI300A eliminates explicit host--device data transfers, the resulting contention for shared memory bandwidth can outweigh this advantage, leading to lower overall performance despite comparable computational capabilities and higher nominal peak memory specifications.

\subsection{Architecture-Driven Performance Divergence}
\label{sec:perf-analysis}

The per-platform analysis above shows that runtime behavior is determined by more than nominal compute capability. On the discrete accelerators, performance broadly follows the expected memory-bandwidth trend: higher effective HBM bandwidth and faster host-device communication reduce the duration of the dominant streaming kernels. The MI300A deviates from this trend because CPU-side SHA3 hashing, FO key derivation, and GPU kernels share the same HBM3 memory subsystem, reducing the effective bandwidth available to the accelerator. Peak bandwidth alone is therefore insufficient to characterize unified-memory architectures; the organization of the memory hierarchy and the overlap between CPU and GPU traffic are equally important.

The traces also expose OpenMP performance-portability limits that are relevant beyond this application. NVIDIA platforms employ the proprietary NVHPC PTX backend, whereas AMD platforms rely on the LLVM OpenMP offload path targeting the \texttt{gfx942} ISA. Consequently, OpenMP constructs such as \texttt{teams distribute parallel for}, reductions, and \texttt{simd} loops are mapped through different backend heuristics for occupancy, scheduling, and register use. Moreover the RNG integration shows that source-level OpenMP portability does not automatically cover all behavior of external device libraries: allocations internal to the HIP RNG adapter can still introduce approximately 30--50\,ms of additional overhead per iteration on AMD systems. 
Overall, OpenMP Target provides effective source-level portability across GPU vendors, but runtime performance remains dependent on memory-system organization, interconnect efficiency, compiler maturity, and the behavior of interoperating device libraries.

\section{Conclusion and Future Work}
\label{sec:concl}
In this work, we present a fully GPU-offloaded implementation of a lattice-based post-quantum KEM built upon the LWE problem. While our previous work primarily focused on comparing CPU and GPU execution on NVIDIA accelerators through OpenACC-based parallelization~\cite{liberati}, this study investigates the feasibility of a fully portable OpenMP Target offloading strategy capable of executing across heterogeneous GPU architectures, including AMD accelerators.

A key contribution of this work is the transition from a CUDA-oriented implementation model toward a portable OpenMP-based workflow. One of the main technical challenges is the integration of the \texttt{RNGonGPU} library, which was originally designed around CUDA-specific assumptions and NVIDIA runtime semantics. Supporting AMD accelerators requires hipifying the RNG backend and exposing OpenMP-managed device memory through HIP interoperability mechanisms, enabling cryptographically secure random number generation to remain fully GPU-resident across both NVIDIA and AMD systems.

The experimental results demonstrate that OpenMP Target offloading can effectively accelerate memory-intensive post-quantum cryptographic workloads while preserving portability across heterogeneous architectures. As shown in Table~\ref{tab:performance_openomp_target}, GPU execution achieves speedups exceeding two orders of magnitude over the multicore OpenMP CPU implementation for large lattice dimensions. The cross-platform comparison reported in Figure~\ref{fig:gpu_perf} and Table~\ref{tab:gpu_relative_slowdown} further shows that NVIDIA GH200 achieves the best overall performance, while AMD MI300X follows closely with only marginal slowdown relative to the GH200 baseline. These results are consistent with the high-bandwidth HBM3/HBM3e memory subsystems available on both platforms, which efficiently sustain the streaming memory accesses dominating the encapsulation and decapsulation kernels.

In contrast, NVIDIA A100 exhibits moderately longer execution times due to the lower effective bandwidth of its HBM2e subsystem and the overhead associated with its PCIe-based communication. The AMD MI300A instead exposes an interesting unified-memory behavior. Despite relying on HBM3 memory comparable to MI300X, the MI300A achieves the weakest scalability among the evaluated systems. Profiling analysis indicates that this degradation originates from contention effects inside the shared CPU--GPU memory subsystem, where CPU-side SHA3 hashing and FO computations compete with GPU kernels for the same HBM3 resources. Compared with the MI300X, the duration of the dominant memory-intensive phases increases by approximately 38\%, highlighting the impact of CPU--GPU bandwidth contention on this workload. These observations demonstrate that performance portability for memory-bound post-quantum cryptographic workloads depends not only on raw compute capability, but also on memory-system organization, runtime maturity, and host--device communication mechanisms.


These results highlight the importance of portable acceleration strategies for memory-bound post-quantum cryptographic workloads. The computational overhead introduced by PQC schemes remains one of the primary barriers to their large-scale deployment in practical systems. Portable GPU programming models such as OpenMP Target can substantially reduce these overheads while avoiding vendor lock-in and architecture-specific implementations, thereby facilitating the transition from classical cryptographic protocols vulnerable to quantum attacks toward deployable quantum-resistant infrastructures.
At the same time, our analysis demonstrates that the performance of LWE-based PQC KEMs is fundamentally constrained by the memory subsystem rather than computational throughput. Since these workloads are dominated by streaming accesses to the public matrix $A$ and random number generation, memory bandwidth and data locality have a greater impact on performance than raw compute capability.
The MI300A case study further illustrates this point. Although its unified CPU--GPU memory architecture eliminates explicit host--device transfers, it also introduces contention when CPU-side operations, such as SHA3 hashing and the FO transform, execute concurrently with GPU kernels and compete for the same HBM3 memory bandwidth. Consequently, the reduction in effective memory bandwidth outweighs the benefits of eliminating data transfers, emphasizing that efficient memory-system design is a key requirement for scalable GPU acceleration of post-quantum cryptography.

To support reproducibility and future research, the complete implementation together with both the OpenACC and OpenMP Target versions is publicly available in the project repository\hyperlink{fn:myfoot}{\textsuperscript{7}}.
Future work will focus on extending the proposed optimization strategy to standardized lattice-based cryptographic schemes, improving runtime efficiency on unified-memory systems, and investigating scalability across multi-GPU and distributed-memory environments.

\begin{credits}
\subsubsection{\discintname}
The authors have no competing interests to declare that are relevant to the content of this article. T.L. developed the OpenMP-parallelized CPU implementation. T.L. and N.S. contributed equally to the OpenMP offloading of the application. N.S. performed the HIPIFY porting of the \texttt{RNGonAMDGPU} library. T.L., N.S., S.R., and E.B. conducted the experimental campaign. E.B. carried out the energy analysis. M.P. conceived the study. All authors contributed to the interpretation of the results and to the writing and revision of the manuscript.

During the preparation of this article, artificial intelligence tools (ChatGPT and Grammarly) were used exclusively to assist with writing and language editing. All scientific content and analyses are the responsibility of the authors.
\end{credits}

\nocite{*}
\bibliographystyle{splncs04}
\bibliography{bib/refs}
\end{document}